\documentclass[aps,prl,reprint,amsmath,amssymb,superscriptaddress]{revtex4-2}

\usepackage{graphicx}
\usepackage{bm}
\usepackage{dcolumn}
\usepackage{xcolor}

\begin{document}

\title{Practically universal representation of the Helfand-Werthamer\\ upper critical field for any transport scattering rate} 

\author{Ruslan Prozorov}
\affiliation{Ames Laboratory, Ames 50011, IA, U.S.A.}
\affiliation{Department of Physics \& Astronomy, Iowa State University, Ames 50011,	U.S.A.}

\author{Vladimir G. Kogan}
\affiliation{Ames Laboratory, Ames 50011, IA, U.S.A.}

\date{20 July 2024} 

\begin{abstract}
The simplified scaling of the orbital upper critical field, $H_{c2}(T)$, for the isotropic case is discussed. To facilitate the analysis of the experimental data, we suggest a simple but accurate approximation in the entire temperature range of the scaled upper critical field valid for any transport scattering rate $H_{c2}/H_{c2}(0) \approx (1-t^2)/(1+0.42 \,t^{1.47})$.
\end{abstract}

\maketitle

\section{Introduction}

The upper critical field, $H_{c2}$, is one of the fundamental parameters characterizing a superconductor. Its temperature dependence is linked to the anisotropies of the Fermi surface and the order parameter. It is also sensitive to magnetic (spin-flip) and non-magnetic (potential or transport) scattering as well as paramagnetic enhancement of the internal magnetic field and spin-orbit scattering. Analysis of $H_{c2}\left(T\right)$ can provide important information about these parameters. 

The evaluation of $H_{c2}\left(T,\rho\right)$ for isotropic $s-$wave superconductor with arbitrary transport (non-magnetic) scattering rate,
\begin{equation}
 \rho=\frac{\hbar}{2\pi T_c\tau}\,,
 \label{rho}
\end{equation}
\noindent where $\tau$ is the mean scattering time, was derived by E. Helfand and N. R. Werthamer (HW) \cite{HW}. The transport dirty limit was discussed by De Gennes \cite{DeGennes} and Maki \cite{Maki}. 

\begin{figure}[tb]
\includegraphics[width=8cm] {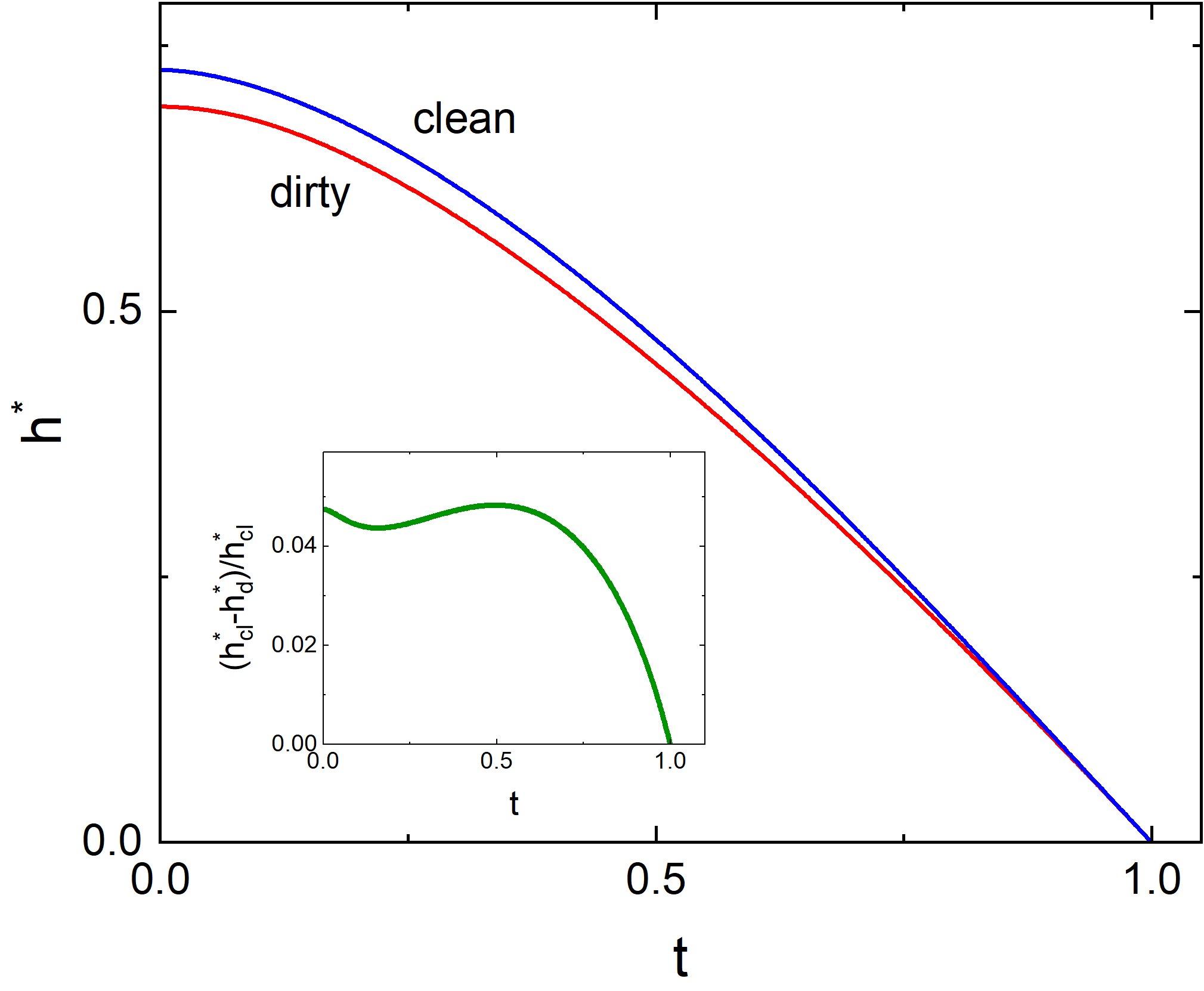}
\caption{
The HW scaling function $h^*(t) = -H_{c2}(t) / (dH_{c2}/dt)_{t\to 1}$ for the clean limit, $h^*_{cl}$ and the dirty limit, $h^*_{d}$, vs. the reduced temperature $t=T/T_c$. Inset: the relative difference between these two curves, $(h^*_{cl}-h^*_{d})/h^*_{cl}$, showing a difference close to 5\% in most of the temperature range, $t<0.7$.
}
\label{f1}
\end{figure}

In their seminal work \cite{HW}, HW have shown that the upper critical field normalized on its slope at $T_c$,
\begin{equation}
 h^*=-\frac{H_{c2}}{T_c (dH_{c2}/dT)_{T_c} }  
 \label{h*}
\end{equation}
\noindent only weakly depends on the scattering rate, $\rho$, as shown in Fig.\,\ref{f1}. One can say that $h^*(t,\rho)$ is close to universal for any $\rho$. It reaches zero-temperature values, $h^*(0)=7\zeta(3)e^{2-C}/48\approx 0.7273$ in the clean case $\rho=0$ and $h^*(0)=(\pi^2/8) e^{-C}\approx 0.6927$ in the dirty limit $\rho>>1$ ($C\approx0.5772$ is the Euler constant). Hence, the relative difference between the clean and dirty values of  $h^*(T)$ is less than $5\%$ at $T=0$, decreases to zero at $T_c$, but is still on a $2\%$ level at $t=T/T_c=0.9$. This is illustrated in the inset in Fig.\,\ref{f1}. 

The ``near universality" of $h^*$ is a useful property commonly used in data analysis. In this note, we point out that a different normalization, namely $u(t) = H_{c2}(T,\rho) /H_{c2} (0,\rho)$, practically collapses the upper critical field curves onto one, much closer to a truly universal curve, compared to the original HW normalization. Such a degree of scaling behavior is sufficient to provide a simple formula for this practically universal curve, which may help the researchers to analyze the experimental data without worrying about the level of disorder always present in real samples.   

The practical usefulness of this scaling representation of the upper critical field was recognized before. For example, Sun and Maki used it to compare theoretical $H_{c2}(T)$ in different directions in $p-$wave superconductors \cite{Sun1993}. Godeke \emph{at al.} used it to show on one graph the experimental data of a variety of 27 Nb-Sn based superconductors and discussed the results in terms of the dirty-limit Maki - de Gennes model (our Eq.~(\ref {h-d})) \cite{Godeke}. However, to the best of our knowledge, this note is the first in which such scaling behavior is used to address arbitrary transport scattering. This is practically relevant since the exact evaluations are time-consuming and, as it turns out, they do  not add new features to the shape of the $H_{c2}(T)$ curve.

\section{The universal curve}

Consider the quantity, 
\begin{equation}
 u(t )=  \frac{H_{c2}(T,\rho)}{H_{c2} (0,\rho)}=  \frac{h (t,\rho)}{h (0,\rho)} \,,
 \label{b}
\end{equation}
\noindent where the reduced field 
\begin{equation}
 h=\frac{H_{c2}}{\phi_0}\frac{\hbar^2 v^2}{2\pi  T_c^2 }\,. 
 \label{h1}
\end{equation}

Clearly, $u(0)=1$ is independent of the scattering rate. To see how this quantity behaves with temperature, let us consider separately the clean and dirty limits near $T_c$. We denote the clean limit by a ``cl" subscript, and the dirty limit by ``d" subscript.

The general expression for the slope $(dh/dt)$ at $T_c$ is given by HW \cite{HW}:
\begin{equation}
- \frac{dh}{dt} \Big |_{t=1} =3\rho^2 \left[\psi\left(\frac{1}{2}\right)-\psi\left(\frac{1+\rho}{2}\right)+\frac{\pi^2\rho}{4}\right]^{-1}.\qquad
\label{slope}
\end{equation}
In the clean case this gives  
\begin{equation}
 -\frac{dh_{cl}}{dt}\Big |_{t=1}  = - \frac{24}{\psi^{\prime\prime} ( 1/2)} =\frac{12}{7\zeta(3)}\,.
\label{slope-clean}
\end{equation}
According to Eq.\,(\ref{b}) the slope $u^\prime (t)=h^\prime (t)/h(0) $. Further, the clean case $H_{c2}(0)=\pi\phi_0T_c^2e^{2-C}/2\hbar^2v^2$ (see e.g. \cite{KP1}), which gives the reduced value,
 \begin{equation}
 h_{cl}(0) = \frac{e^{2-C}}{ 4}  \,.
 \label{hcl(0)}
\end{equation}
Hence, we obtain
 \begin{equation}
 -\frac{du_{cl}}{dt}\Big |_{t=1}  =  \frac{48\,e^{ C-2}}{7\zeta(3)} \approx 1.3750\,.
 \label{b-slope-clean}
\end{equation}
Note that this number is just the inverse of the HW clean-limit number $h^*_{cl}(0)=0.727$, because in the $h^*$ normalization, the slope at $T_c$ is 1.

\begin{figure}[tb]
\includegraphics[width=8cm] {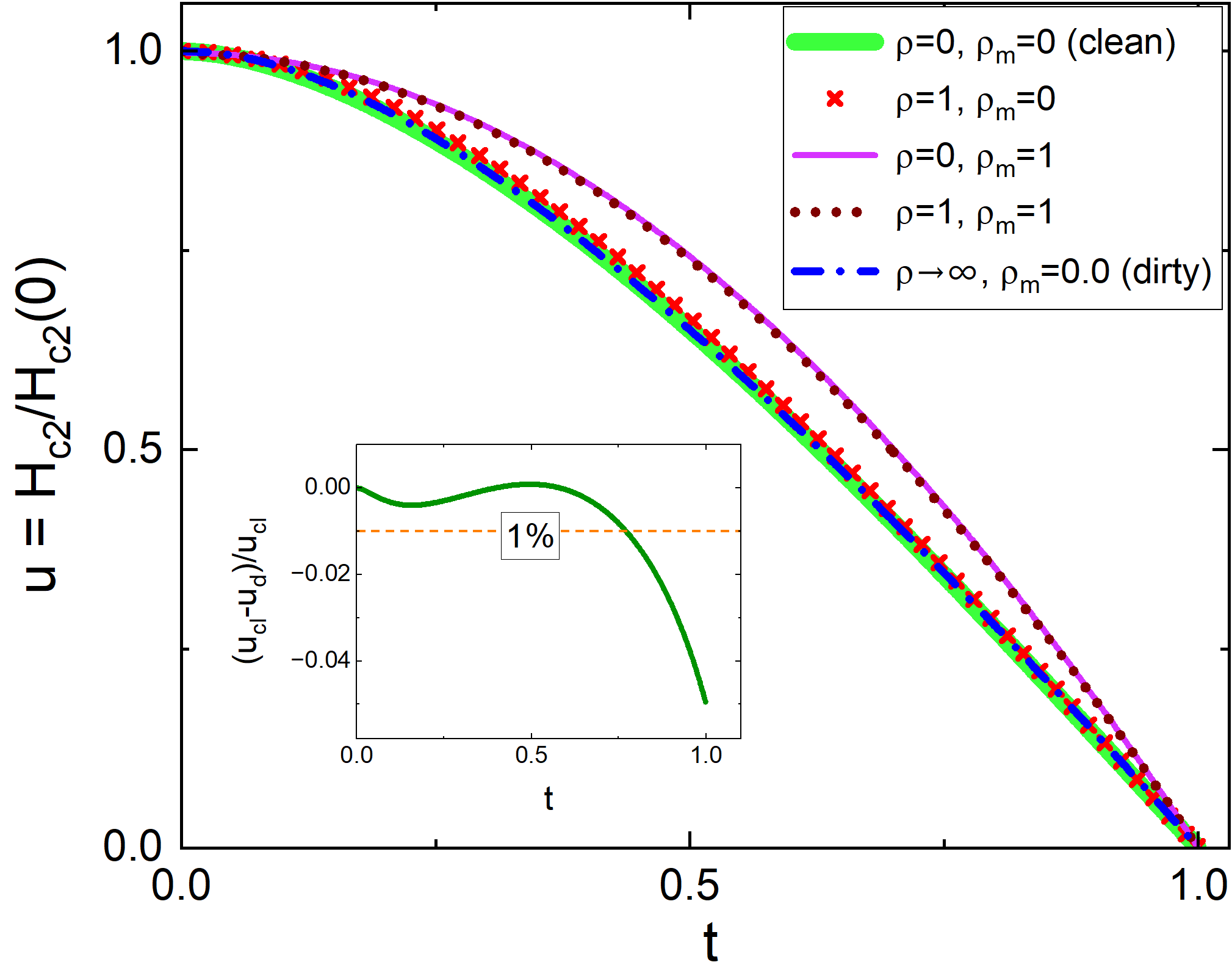}
\caption{
The scaled function $u(t)=h(t)/h(0)$ for different types of scattering. The wide solid green line shows the numerical result in the clean case; the blue dash-dotted line is the transport dirty limit; the red x-marks show the finite transport scattering rate case, $\rho=1$, $\rho_m=01$. These three curves are practically collapsed on a single universal curve. In stark contrast, magnetic scattering results in a more concave curve, shown for pure magnetic scattering, $\rho_m=1$ and $\rho=0$ (violet solid line), and for a more realistic mixed case, $\rho_m=1$ and $\rho=1$ (blue dotted line). Inset: the relative difference, $(u_{cl}-u_{d})/u_{cl}$, between the two limiting cases showing about 0.5\% difference level for most of the temperature interval.
}
\label{f2}
\end{figure}

In the dirty limit, Eq.\,(\ref{slope}) yields
\begin{equation}
 - \frac{dh_{d}}{dt}\Big |_{t=1} =   \frac{12\rho}{\pi^2}  \,.
 \label{slope-dirty}
\end{equation}
The dirty limit self-consistency equation \cite{DeGennes,Maki}, Eq.(40) of the HW paper \cite{HW},
\begin{equation}
-\ln t =  \psi\left(\frac{1}{2}+\frac{h}{6\rho t}\right)- \psi\left(\frac{1}{2} \right)\, 
\label{h-d}
\end{equation}
yields  
\begin{equation}
h_d(0)=   \frac{3\rho}{2\,e^C}\,. 
\label{h-d(0)}
\end{equation}
Thus, we have, 
\begin{equation}
\frac{du_{d}}{dt}\Big |_{t=1}  =  \frac{8\,e^{ C}}{\pi^2} \approx 1.4437 \,.
\label{b-slope-dirty}
\end{equation}
Note that  this number is just the inverse of the HW number $h^*_d(0)=0.693$. 

\begin{figure}[tb]
\includegraphics[width=8cm] {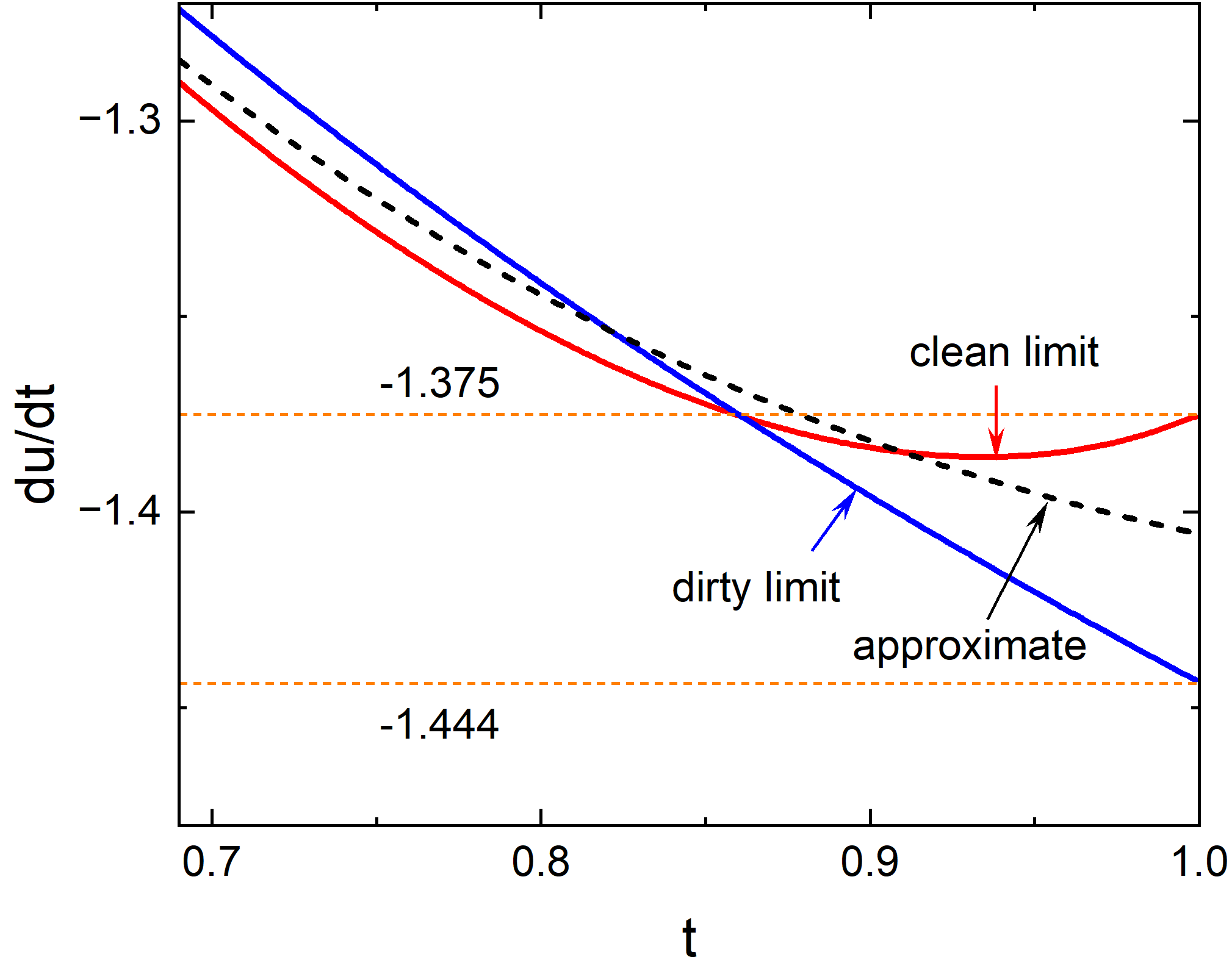}
\caption{The slopes $du/dt$ near $T_c$. The solid red curve shows the clean limit, the solid blue curve shows the dirty limit, and the dashed black line shows the derivative of the suggested universal curve, Eq.~(\ref{U1(t)}). The dashed orange lines with numbers show the analytic limits, Eq.~(\ref{b-slope-clean}) and Eq.~(\ref{b-slope-dirty}). The suggested universal approximation, based on Eq.~(\ref{U1(t)}), fits in between the limiting values.}
\label{f3}
\end{figure}

Numerically, one can evaluate $H_{c2}(t,\rho)$ at any temperature using either the original HW equations or a revised theory \cite{KP1} better suited for numerical analysis to obtain $u(t)$ for clean, dirty (and any intermediate) cases. The latter approach can also be used to compute $H_{c2}$ in the case of finite magnetic and non-magnetic scattering. The wide solid green line in Fig.\,\ref{f2} shows the numerical result in the clean case; the blue dash-dotted line shows the transport dirty limit; the red x marks show the case of borderline transport scattering, $\rho=1$, when the coherence length $\xi$ and the mean free path, $\ell$, are of the same order. These three curves practically collapse on a single universal curve and, as shown later, can be well approximated by a simple equation $u(t)\approx (1-t^2)/(1+0.42 \,t^{1.47})$. Note that the magnetic scattering results in a concave deviation from the universal curve. The solid violet line in Fig.,\ref{f2} shows the case of pure magnetic scattering, $\rho_m=1$ and $\rho=0$, and the blue dotted line shows a mixed case, $\rho_m=1$ and $\rho=1$. This shows that the pair-breaking scattering will manifest itself in this way regardless of the background transport disorder, which makes such a simple analysis very useful.
The inset in Fig.\,\ref{f2} shows the relative difference, $(u_{cl}-u_{d})/u_{cl}$, which is just a downward shift of a similar quantity shown in the inset of Fig.,\ref{f1}, by approximately 0.048, resulting in about ten times smaller relative difference in most of the temperature range.
 
Thus, $u(t)$ is a nearly universal function of $t$. Inspired by the Casimir and Gorter theory, see Eq.(9) in Ref.\cite{Doug}, we used the following form to fit the ``exact" numerical results,
\begin{equation}
u(t) = \frac{1-t^2}{1+a t^b}\,,
 \label{U(t)}
\end{equation}
\noindent  The fitting yields $a=0.423,\;b=1.468$ in the clean case and $a=0.416,\; b=1.472$ in the dirty limit. The results are so close, that we converged on a single function that describes the scaled $u(t)$ curve for any transport scattering rate with a good accuracy,
\begin{equation}
u(t) \approx \frac{1-t^2}{1+0.42 t^{1.47}}\,,
 \label{U1(t)}
\end{equation}
It is worth noting again that, as shown in the inset to Fig.\,\ref{f2}, the ``universality parameter" $(u_{cl}-u_{d})/u_{cl}$ for this new scaling is an order of magnitude smaller than that for original HW, compare with the inset in Fig.\,\ref{f1}.

\begin{figure}[tb]
\includegraphics[width=8cm] {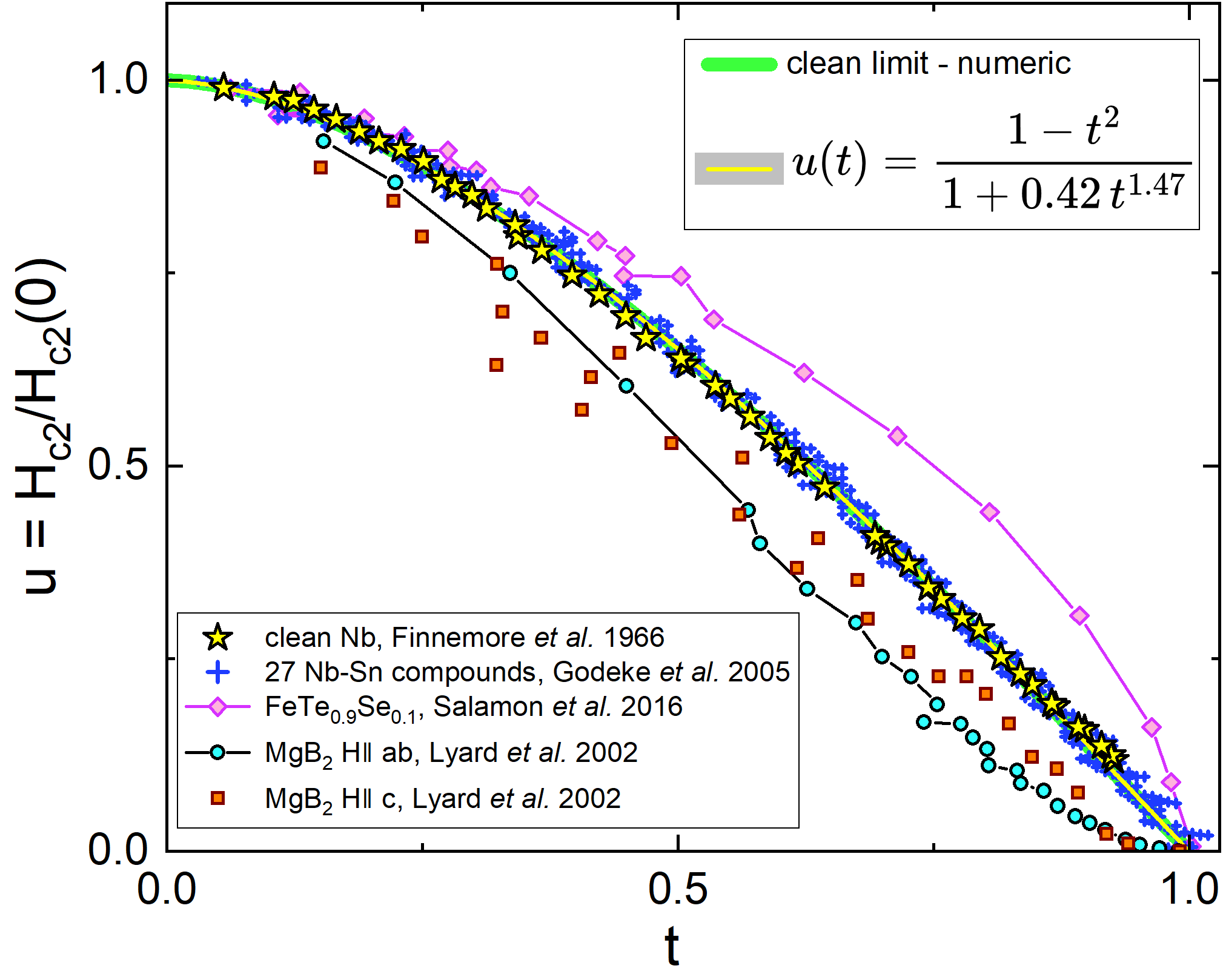}
\caption{
Illustration of the use of the suggested scaling for the analysis of the experimental data. The wide green line shows the numerical clean limit and solid yellow line shows the universal scaling function, Eq.~(\ref{U1(t)}). The data that follow this scaling very well belong to cubic $s-$wave materials, represented here by Nb-Sn - based superconductors that include 16 ternary and 11 binary compounds \cite{Godeke} and a very clean niobium, shown by stars \cite{Doug}. This contrasts with the data shown by violet rhombi for FeTe$_{0.9}$Se$_{0.1}$ films that had known magnetic scattering \cite{Salamon}. It exhibits a concave curvature, consistent with Fig.\ref{f2}, where magnetic scattering is considered. Another case, leading to the convex $u(t)$ curve, is a two-gap $s_{++}$ superconductor, MgB$_2$, shown by cyan-filled black circles for $H\parallel ab-$plane and by orange-filled black squares for the $H\parallel c-$axis \cite{Lyard}.
}
\label{f4}
\end{figure}
 
To investigate the source of some non-universality of $u$ close to $T_c$, we numerically calculated the slopes $du/dt$. Figure\,\ref{f3} shows that the slopes, $du/dt$, near $T_c$ differ for the clean and dirty cases, approaching $-1.375$ at $t=1$ in the clean case, Eq.~(\ref{b-slope-clean}) and $-1.444$ in the dirty limit, Eq.~(\ref{b-slope-dirty}). However, this relatively small difference does not show in the plot of $u(t)$ since near $T_c$, $u\to 0$. For comparison, the dashed black line in Fig.\,\ref{f3} shows the temperature derivative of the suggested universal curve, Eq.~(\ref{U1(t)}). Expectedly, it falls in between the limiting slopes.

To illustrate the practical usability of our results, Fig.\ref{f4} shows several cases where $H_{c2}(T)$ agrees with our ``universal" curve, Eq.~(\ref{U1(t)}), or not for the reasons provided. Among many reports of the upper critical field, we sought the most complete data covering the entire temperature range. The reference lines in Fig.\ref{f4} are: the wide green line shows the numerical clean limit (almost invisible under data points), and the solid yellow line shows a simplified universal $u(t)$ of Eq.~(\ref{U1(t)}). The data that follow this scaling very well belong to cubic $s-$wave materials, represented here by Nb-Sn - based superconductors that include a rich compilation of 16 ternary and 11 binary compounds \cite{Godeke}. We note that it is likely that these 27 compounds had different levels of scattering, yet they all collapsed on a single universal curve. To reinforce this argument, stars in Fig.\ref{f4} show a very clean niobium metal \cite{Doug}. This is contrasted with FeTe$_{0.9}$Se$_{0.1}$ data, shown by violet rhombi, which had known magnetic scattering \cite{Salamon} and exhibits a concave curvature, consistent with Fig.\ref{f2}. Another case, leading to the convex-type deviation, is the well-known two-gap superconductor, MgB$_2$, shown by cyan-filled black circles for $H\parallel ab-$plane and by orange-filled black squares for the $H\parallel c-$axis \cite{Lyard}. 

The proposed scaling approach is naturally justified considering that the error bar in the experimental determination of $H_{c2}$ is usually not small, so the precision of the proposed scaling function, Eq.(\ref{U1(t)}), is quite sufficient as can be seen in Fig.\ref{f4} where real-world experimental data are shown.

\section{Conclusions}
In conclusion, we show that the analysis of the upper critical field in scaled form $u(t)=H_{c2}(T/T_c)/H_{c2}(0)$ provides a simple way to check the compliance of the experimental results with the HW orbital critical field. One has to use Eq.~(\ref{U1(t)}) with a two-parameter fit, $H_{c2}(0)$ and $T_c$, to scale the data irrespective of the transport scattering rate. Significant deviations will signal the presence of other mechanisms that influence $H_{c2}(T)$. For example, multi-gap $s_{++}$ superconductivity will show as a convex curve with a noticeable curvature in the upper half of the temperature range, whereas magnetic (possibly pairbreaking in general) scattering will result in a more pronounced concave curve above the universal $u(t)$ line.

\vspace{0.3cm}

\begin{acknowledgments}
R.P. acknowledges the useful discussions with A. Gurevich. This research was supported by the U.S. Department of Energy, Office of Basic Energy Sciences, Division of Materials Sciences and Engineering. Ames National Laboratory is operated for the U.S. Department of Energy by Iowa State University under Contract No.DE-AC02-07CH11358. 
\end{acknowledgments}


\begin{thebibliography}{99}

\bibitem{HW}E. Helfand and N. R. Werthamer, Phys. Rev. {\bf 147}, 288 (1966).

\bibitem{DeGennes}   P. G. de Gennes, Phys. Kondens. Materie {\bf 3}, 79 (1964).

\bibitem{Maki} K. Maki, Physics {\bf 1}, 21 (1964).

\bibitem{Sun1993}Y. Sun, K. Maki, Phys. Rev. B {\bf 47}, 9108 (1993).

\bibitem{Godeke}A. Godeke, M. C. Jewell, C. M. Fischer, A. A. Squitieri, P. J. Lee, and D. C. Larbalestier,  J. Appl. Phys. {\bf 97}, 093909 (2005).

\bibitem{KP1} V. G. Kogan and R. Prozorov, Phys. Rev. B  {\bf 88}, 024503 (2013).

\bibitem{Doug} D. K. Finnemore, T. F. Stromberg, C. A. Swenson, Phys. Rev. {\bf 149}, 231 (1966).

\bibitem{Salamon} M. B. Salamon, N. Cornell, M. Jaime, F. F. Balakirev, A. Zakhidov, J. Huang, H. Wang, Scientific Reports {\bf 6}, 21469 (2016).

\bibitem{Lyard} L. Lyard, P. Samuely, P. Szabo, T. Klein, C. Marcenat, L. Paulius, K. H. P. Kim, C. U. Jung, H.-S. Lee, B. Kang, S. Choi, S.-I. Lee, J. Marcus, S. Blanchard, A. G. M. Jansen, U. Welp, G. Karapetrov, W. K. Kwok, Phys. Rev. B {\bf 66}, 180502 (2002).

\end{thebibliography}
\end{document}